\documentclass[apj]{emulateapj}

\newcommand{\myemailb}{Lauren.MacArthur@nrc-cnrc.gc.ca}
\newcommand{\lt}{\ifmmode\,<\,\else \,$<$\,\fi}
\newcommand{\kms}{\ifmmode\,{\rm km}\,{\rm s}^{-1}\else km$\,$s$^{-1}$\fi}
\newcommand{\sersic}{S\'{e}rsic}

\newcommand{\magarc}{\ifmmode {{{{\rm mag}~{\rm arcsec}}^{-2}}}
             \else {{{mag}$~${arcsec}$^{-2}$}}
             \fi}
\newcommand{\hunit}{km~s$^{-1}$~Mpc$^{-1}$}
\newcommand{\hub}{H$_{\hbox{\scriptsize 0}}$}
\newcommand{\etal}{et~al.\@}
\newcommand{\eg}{e.g.\@}
\newcommand{\ie}{i.e.\@}

\shorttitle{Environmental Evolution of Galactic Bulges}

\begin{document}

\title{Environmental Effects in the Evolution of Galactic Bulges}


\author {Lauren A. MacArthur\altaffilmark{1,2},
Richard S. Ellis\altaffilmark{1},
Tommaso Treu\altaffilmark{3},
Sean Moran\altaffilmark{4}
}

\altaffiltext{1}{Department of Astrophysics, California Institute of 
                Technology, MS 105-24, Pasadena, CA 91125}
\altaffiltext{2}{Herzberg Institute of Astrophysics, National Research 
                Council of Canada, 5071 West Saanich Road, Victoria, BC 
                V8X 4M6, Canada; \myemailb}
\altaffiltext{3}{Department of Physics, University of California,
                 Santa Barbara, CA 93106-9530; tt@physics.ucsb.edu} 
\altaffiltext{4}{Department of Physics \& Astronomy, The Johns Hopkins 
                 University, Baltimore, MD 21218; moran@pha.jhu.edu}


\begin{abstract}
We investigate possible environmental trends in the evolution of
galactic bulges over the redshift range 0\,$<$\,$z$\,$<$\,0.6.  For
this purpose, we construct the Fundamental Plane (FP) for cluster and
field samples at redshifts $\langle$\,$z$\,$\rangle$\,=\,0.4 and
$\langle$\,$z$\,$\rangle$\,=\,0.54 using surface photometry based on
$HST$ imaging and velocity dispersions based on Keck spectroscopy. As
a reference point for our study we include data for pure ellipticals,
which we model as single-component \sersic\ profiles; whereas for
multi-component galaxies we undertake decompositions using
\sersic\ and exponential models for the bulge and disk respectively.
Although the FP for both distant cluster and field samples are offset
from the local relation, consistent with evolutionary trends found in
earlier studies, we detect significant differences in the zero point of
$\simeq$\,0.2\,dex between the field and cluster samples at a given
redshift. For both clusters, the environmentally-dependent offset
is in the sense expected for an accelerated evolution of bulges in
dense environments. By matching the mass range of our samples, we
confirm that this difference does not arise as a result of the
mass-dependent downsizing effects seen in larger field samples. Our
result is also consistent with the hypothesis that -- at fixed mass and
environment -- the star formation histories of galactic bulges and
pure spheroids are indistinguishable, and difficult to reconcile with
the picture whereby the majority of large bulges form primarily via
secular processes within spiral galaxies.
\end{abstract}


\section{Introduction}\label{sec:intro}

A spheroid can exist as an independent structure, \ie\ a pure
elliptical galaxy, or it can be surrounded by a disk component, \ie\
the bulges of spiral and lenticular galaxies. Good progress has been
achieved in the past decade in constraining the past star formation
history of spheroid-dominated E/S0 galaxies through studies of the
redshift-dependent Fundamental Plane (FP) - the relation between
galaxy size, surface brightness, and stellar velocity dispersion (\eg\
Treu \etal\ 2002; Treu \etal\ 2005; van Dokkum \& van der Marel 2007).
More recently, these analyses have been extended to include the bulges
of spiral galaxies (MacArthur \etal\ 2008, hereafter Mac08; Hathi
\etal\ 2009), allowing a direct comparison between the mass assembly
history of pure spheroids and those residing in disks. This enables us
to test the simple conjecture that galactic bulges share a similar
assembly history to isolated spheroids and are not, at least to first
order, significantly influenced in their growth by disk-related
processes. While the bulge samples remain small and can only currently
be compiled to redshifts $z$\,$\lt$\,1, the general picture that has
emerged is that both isolated spheroids and galactic bulges evolved at
a similar rate {\it for a given spheroid mass} and with similar
mass-dependent trends. In particular, for both systems the less
massive examples have witnessed more recent activity whereas the most
massive ones formed the vast majority of their stars at high redshift
($z_f$\,$\gtrsim$\,2).

A key question of interest for both bulges and isolated spheroids is
the role of the local environment. While the influence of the local
environment on the morphological mix (Dressler 1980; Smith
\etal\ 2005), colors (Bamford \etal\ 2009), mass-to-light ratios
(Moran \etal\ 2005; hereafter M05), and star formation rates (Lewis
\etal\ 2002; Kauffmann \etal\ 2004; Cooper \etal\ 2008) of galaxies is
now well-established, some studies have emphasized the importance of
mass as the governing evolutionary factor (\eg\ Treu
\etal\ 2005; Mac08). Separating the effects of mass from that of
environmental density is thus clearly important. Environmental trends
are qualitatively consistent with the predictions of popular structure
formation models where growth is accelerated in more massive halos
which start their collapse at earlier times (Governato \etal\ 2008).
By contrast, mass-dependent trends require feedback processes which
are poorly understood.

For the E/S0 population, the question of the relative importance of
the environment is still somewhat open. In contrast to field versus
cluster trends found in early work (Treu \etal\ 2002), van Dokkum \&
van der Marel (2007) found only marginal differences in the inferred
ages for massive galaxies as a function of their environment. However,
M05, studying E/S0s spanning a wide range of cluster-centric distance
within a single cluster, Cl\,0024+17 ($z$\,=\,0.4), found a
significant radial trend. This is in the sense of a decreasing $M/L$
ratio with both cluster-centric distance and local density, so that
early-types close to the cluster core have older ages, while those at
the periphery are younger, and more similar to field galaxies at a
similar redshift.  A key concern in attempting to reconcile these
various results is the mass range explored in the various samples, as
it is known that the mass function depends on environment.

Clearly, it is desirable to characterize the past history of bulges
and spheroidal galaxies as a function of both mass and
environment. Recognizing this, we extend the scope of our earlier work
(Mac08) and present here a comparative field versus cluster FP
analysis in two redshift bins defined by the clusters Cl\,0024+17
($z$\,=\,0.4) and MS\,0451$-$03 ($z$\,=\,0.54).  FP parameters are
derived from Keck/DEIMOS spectroscopy to measure velocity dispersions
and $HST$ imaging for photometric parameters, supplemented with ground
based CFHT and SUBARU imaging.  For all distance- dependent quantities
we adopt a flat cosmological model with $\Omega_M$\,=\,0.3,
$\Omega_{\Lambda}$\,=\,0.7, and
\hub\,=\,65~\hunit. All magnitudes are in the AB system (Oke 1974).

\section{Data}\label{sec:data}

Imaging data for the field samples is largely taken from the GOODS
survey (Giavalisco \etal\ 2004) which provides deep imaging in four
ACS passbands.  The broad wavelength coverage permits the derivation
of accurate $k$-corrections for the selected galaxies, which all have
measured spectroscopic redshifts from the Keck Team Redshift Survey
(Wirth \etal\ 2004).  Additional field galaxies were included from the
sample of spectroscopically-confirmed non-members located within the
two cluster fields (see Moran \etal\ 2007a).

Imaging data for the cluster samples is available from previous $HST$
campaigns described fully in Moran \etal\ (2007b; hereafter
M07b). Both clusters were observed in the F814W filter with WFPC2 (for
Cl\,0024+17) and ACS (for MS\,0451-03). Galaxies were selected from
the comprehensive spectroscopic survey described by M07b. Accurate
$k$-corrections were derived using optical ($BVRI$) ground-based
imaging from the CFHT and SUBARU telescopes. Due to the inferior
resolution of the ground-based data, aperture magnitudes of radius
0\farcs6 were used for the $k$-corrections.  Previous studies have
shown (Ellis \etal\ 2001; Mac08) that disk contamination is minimal
within such an aperture.

Both field and cluster galaxies were selected morphologically via
visual inspection to a limit of $i_{AB}$\,$\sim$\,21.5.
Details of the classification and its reliability are presented in
Treu \etal\ (2003) and Bundy \etal\ (2005) for the cluster and field
samples, respectively.

The Keck spectroscopic data arises from a number of independent
campaigns.  The bulk of the velocity dispersions in the cluster fields
derives from a $i_{AB}$\,$<$\,21.5 sample discussed by M05/M07b.  This
sample was augmented via an additional observing run in November 2007
dedicated to increasing the cluster samples for this purpose.  Here,
we selected further cluster galaxies with $i_{AB}$\,$<$\,21.5 and a
bulge/total ($B/T$) fraction $>$\,0.2 determined from decompositions
of the $HST$ imaging. The opportunity also enabled us to also enlarge
the observed sample in the GOODS-S field with respect with the GOODS
field samples presented in Mac08. Typical exposure times on this
latest run were 5--6\,hrs with seeing conditions in the range
0.6--1.1\arcsec.

To facilitate the desired comparison between field and cluster, we
defined the GOODS-N/S field samples to be those in the above
compilation within the redshift ranges 0.3\,$<$\,$z$\,$\le$\,0.5 (for
comparison with the Cl0024+17 sample) and 0.5\,$<$\,$z$\,$\le$\,0.7
(for comparison with the MS0451-03 sample). Nine field galaxies were
accepted from the clustered fields lying within these relevant
boundaries and 4\,$\sigma$ outside the cluster redshift distribution.

The resulting sample comprises 240 spheroids (133 cluster, 107 field)
of which 175 (94 cluster, 81 field) represent bulges in two-component
galaxies and 65 (39 cluster, 26 field) represent single-component
spheroidal galaxies.

\section{Analysis}\label{sec:analysis}

The derivation of FP parameters from the spectroscopic and photometric
data discussed above closely followed the procedures discussed in
detail in Mac08 to which the reader is referred. Briefly, size and
surface brightness parameters were determined from one-dimensional
profiles (see MacArthur, Courteau, \& Holtzman 2003 for details)
following decomposition of the $HST$ images into \sersic\ bulge and
exponential disk components.  For pure elliptical galaxies, only a
single \sersic\ profile was fit. Stellar velocity dispersions were
measured from the Keck/DEIMOS spectra using the Gauss-Hermite Pixel
Fitting algorithm (van~der~Marel 1994).
By limiting the sample to systems with $B/T$\,$>$\,0.2, contamination
from the disk to the central velocity dispersion measurement is
insignificant (Mac08).

While all spectroscopic data sets were obtained with Keck/DEIMOS,
there were a few differences in the observational set-up and
measurement procedures across the various cluster sub-samples which
could affect the field-cluster comparison. These differences reflect
the fact that the present bulge comparison was not envisaged when the
M05/M07b cluster study was planned (and do not apply to the field
samples of Mac08). Two differences are worth considering.

Firstly, the bulk of the early cluster galaxy spectra were taken over
restframe 3500--6700\,\AA\ with a 600\,l/mm grating suitable for E/S0
galaxies, whereas the November 2007 data sampled restframe
3600--5400\,\AA\ with a 1200\,l/mm grating appropriate for less
massive systems. In probing a mixed stellar population, a redder
wavelength range might be more sensitive to older stars.

A second difference arises in the extraction of the spectra.  All data
sets were reduced using the DEEP2 pipeline (Davis \etal\ 2003) which
extracts both a 1-D optimally weighted spectrum, as well as a 2-D
spectrum.  The M05/M07b analyses used the 1-D spectrum to derive
dispersions whereas, for the bulges, only the central bin 
was used with no co-addition of pixels.  The latter data thus have a
fixed effective aperture of 0\farcs35, while the Moran \etal\ data
have a larger, and slightly variable aperture depending on how many
pixels were coadded for the 1-D spectrum.  Conceivably these aperture
differences could lead to subtle biases in our desired comparison.

Fortunately, repeat observations of a number of galaxies between the
two samples allows us to constrain this possible bias.  For
Cl\,0024+17, after correcting to a common effective aperture of radius
$r_{\rm e}/8$, there is no significant average offset for the 19
galaxies common to both data sets 0.01\,$\pm$0.02\,dex.  However, for
MS\,0451-03, after aperture correction, a difference of
0.08\,$\pm$\,0.02\,dex towards larger dispersions for the Moran \etal\
measurements was found.  Although still only a small effect compared
to the final differences that we measure between cluster and field, to
verify the source of this offset we reanalyzed the Moran
\etal\ data using exactly the technique adopted for the new data. We
found that the small offset can, in fact, be attributed to the
different wavelength range. In the re-analysis of the Moran \etal\
data the red end was masked during the fit, producing consistent
results. We conclude that the effect is due to the presence of
composite stellar populations and can be mitigated by focusing on the
same exact wavelength region. Thus, to correct the Moran \etal\
dispersions in MS\,0451-03 to the same scale as the other data, a
shift of 0.08\,dex was added to them for the current analysis. We
stress that this in no way invalidates the earlier discussions of this
data which was internally consistent.

Finally, in order to measure the evolutionary trends, a suitable local
reference FP is required.  For this we use the relation of
J{\o}rgensen, Franx, \& Kjaergaard (1996, hereafter J96) for
early-type Coma cluster galaxies.  This choice does not provide an
ideal comparison for the current study as it does not account for
structural non-homology or any disk component in their sample
galaxies.  The effects of this difference were explored in Mac08,
revealing an offset (equivalent to 0.16\,dex in the local zero point)
in the structural parameters from the fixed $n$\,=\,4 single \sersic\
fits versus the best-fit \sersic\ $n$ having also fit a disk component
when present. Thus, for a direct comparison with our higher-$z$
samples, we adopt the shifted J96 FP as our local zero point.

\begin{figure}
\begin{center}
\plotone{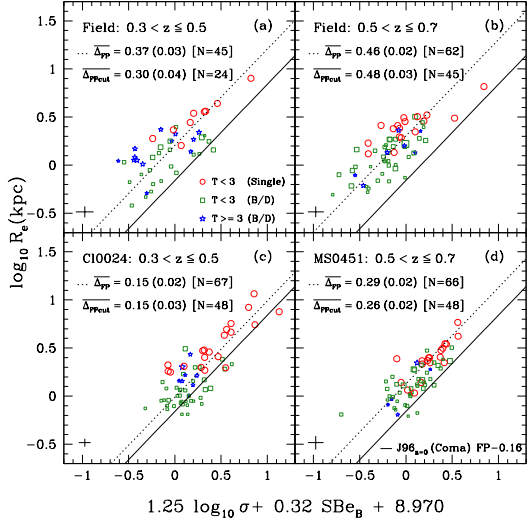}
\caption{The Fundamental Plane (FP) for field (upper panels) and 
         cluster (lower panels) samples in two redshift bins.  Point
         types and colors are as follows: red circles:
         single-component spheroids (\ie\ ``pure'' ellipticals), green
         squares: bulges of two-component systems with
         T-type\,$<$\,3($\equiv$\,Sa+b), and blue stars: bulges of
         two-component systems with T-type\,$\ge$\,3
         and $B/T$\,$>$\,0.2.  The solid lines are the local J96
         relation for Coma corrected as discussed in the text.  Dotted
         lines represent the mean offset from the local FP for each
         sub-sample.  The mean offset, $\overline{\Delta_{{\rm FP}}}$,
         error on the mean, and number of galaxies are indicated at
         the top left of each panel.  The corresponding numbers are
         also given for the mass limited subsamples
         ($\overline{\Delta_{{\rm FPcut}}}$, see
         text). \label{fig:FP}}
\end{center}
\end{figure}
\begin{figure}
\begin{center}
\plotone{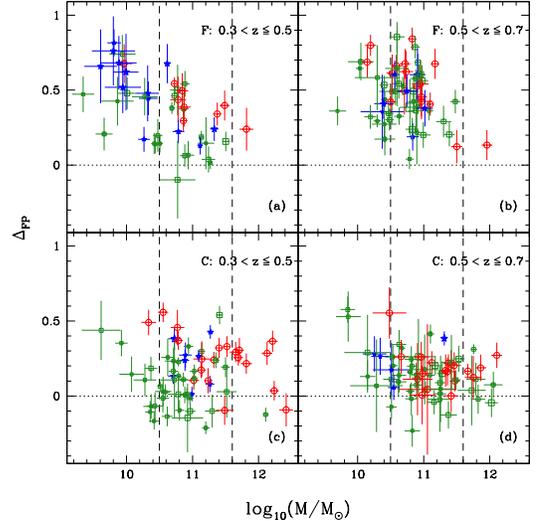}
\caption{Offset from the local FP relation as a function of dynamical 
         spheroid mass.  Point types and colors are as in
         Fig.\@~\ref{fig:FP}.  The dashed vertical lines indicate the
         mass limits appropriate for the restricted sample for the
         $\Delta_{FPcut}$
         (10.5\,$\le$\,log($M/M_{\odot}$)\,$\le$\,11.6) values in
         Fig.\@~\ref{fig:FP}.  \label{fig:DFPM}}
\end{center}
\end{figure}

\section{Results}

Having established homogeneous data sets of field and cluster
spheroids in two redshift bins and a suitable local comparison
relation, we make a direct differential comparison of the FP in each
bin.  In Fig.\@~\ref{fig:FP} we plot the FP for our sample separated
into two redshift bins for the field (top) and cluster (bottom)
samples.  The solid line marks the local relation.  The comparison of
the field versus cluster samples shows a similar difference for the
spheroidal FP zero points of 0.22\,$\pm$\,0.04\,dex and
0.17\,$\pm$\,0.03\,dex for the successive redshift bins, in the sense
that the cluster spheroidals are systematically fainter at fixed size
and velocity dispersion, consistent with older stellar populations.

The key bias that could affect the intended analysis is the mass range
sampled in each redshift bin.  Given our earlier discussion, to
isolate environmental effects from those arising as a result of
mass-dependent effects (Bundy \etal\ 2005; Treu \etal\ 2005;
van~der~Wel \etal\ 2005; Mac08), it is important to consider carefully
the mass range for field and cluster spheroids.  We determine the
dynamical mass using the relation
$M$\,$\equiv$\,$k(n)\sigma^{2}R_e/G$, where $k(n)$ is a profile shape
dependent virial coefficient taken from Trujillo \etal
(2004). Although such a derived mass is affected by the contribution
of the disk and dark matter components, so long as we undertake a
differential comparison, we consider this a minor effect.

Fig.\@~\ref{fig:DFPM} shows the dependence of the FP offset derived
from Fig.\@~\ref{fig:FP} as a function of spheroid mass.  Here we see
the field samples are indeed more heavily weighted towards lower
masses and also show evidence of the downsizing trends discussed, \eg\
by Treu \etal\ (2005). To determine rigorously whether there is an
additional environmental trend, we thus recompute the FP offsets
restricting both cluster and field samples to lie within the mass
interval log($M/M_{\odot}$)\,=\,$10.5$--$11.6$ (chosen such that both
datasets adequately sample the range)\footnote{A K-S test confirms the
mass distribution is not significantly different for the low-$z$ bin,
and examination of selection limits as in Fig.\@~11 in Mac08 confirms
no galaxies in the low-$z$ bin would have been missed in the high-$z$
bin.  Additionally, restricting the field bin to as narrow as
0.53\,$<$\,$z$\,$\le$\,0.56 again does not alter our results.}.  The
new offsets are shown in Fig.\@~\ref{fig:FP} as ``$\Delta_{{\rm
FPcut}}$''. With these refined samples, the new FP zero point shifts
become 0.15\,$\pm$\,0.05\,dex and 0.22\,$\pm$\,0.04\,dex for the
successive redshift bins.  Further restriction on the mass range such
that the sampling in the field and cluster bins are matched also does
not change the results.  In summary, there is a clear environmental
effect even when the mass and $z$ ranges of the field and cluster spheroid
samples are restricted to be the same (\ie\ the trend is not driven
by selection effects).

\section{Discussion}\label{sec:discuss}

\begin{figure}
\begin{center}
\vskip -2.2cm
\plotone{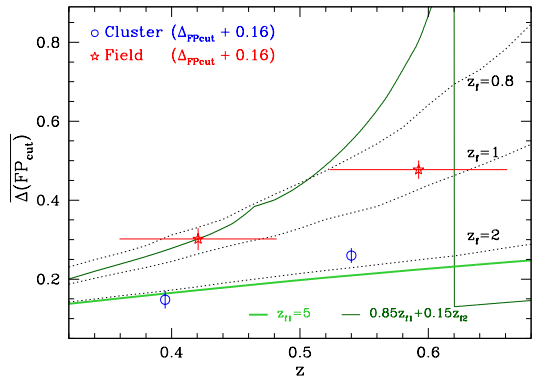}
\caption{Average offset from the local FP relation as a function of redshift
         for the mass-restricted sample for field (red stars) and
         cluster (open circles) samples.  The points for the field
         sample are located at the mean $z$ of the sample and the
         horizontal bars represent the rms spread.  Dotted black lines
         represent model tracks of a passively-evolving single burst
         population of solar metallicity for three formation redshifts
         ($z_f$\,=\,$0.8,1,2$) from the models of Bruzual \& Charlot
         (2003).  The solid lines represent an initial burst of SF at
         $z_{f1}$\,=\,5 (light green) with a second burst comprising
         15\% of the stars by mass added at $z_{f2}$\,=\,0.62 (dark
         green).  \label{fig:DFPz}}

\end{center}
\end{figure}

Fig.\@~\ref{fig:DFPz} summarizes the primary result of this paper.  We
observe a strong environmental signal in both redshift bins for all
spheroids with $B/T$\,$>$\,0.2. This signal is not a manifestation of
mass-dependent trends and thus is in addition to the downsizing trends
observed before. In terms of star formation history, the observed
trends can be interpreted as follows. In the cluster environment, the
spheroidal component of galaxies is consistent with having formed very
few stars below redshift $z$\,$\sim$\,2, in agreement with the general trend
observed for pure spheroidals of the same mass. In contrast, in the
field environment, the evolution of the FP is significantly faster,
consistent with a more recent stellar population (a single star
formation episode would imply formation redshifts of 
$z$\,$\simeq$\,0.8). However -- as in the case of pure spheroidals
(\eg\ Treu \etal\ 2005) -- a more likely interpretation of the
observed trends is that the majority of the stars are formed at
significantly higher redshift, and the integrated stellar populations
are then rejuvenated by secondary episodes of star formation below
$z$\,$\sim$\,1. As an illustration, we show the expected evolution for a
model where 15\% of the stellar mass is formed at $z$\,=\,0.62. The
similarity of the star formation histories of galactic bulges and pure
spheroids appears difficult to reconcile with the picture whereby the
majority of large bulges form primarily via secular processes within
spiral galaxies.

In the case of pure spheroidals of high mass, a decisive argument in
favor of the secondary bursts model is the relatively slow evolution
of their stellar mass function since $z$\,$\sim$\,1 (Bundy \etal\
2005).  Unfortunately, no such measurement is currently available for
the spheroid and stellar components of galaxies since $z$\,$\sim$\,1
to break the degeneracy between star formation history and assembly
history. The recent growth of the spheroidal component of lenticular
galaxies (\eg\, Geach \etal\ 2009) and the observed evolution of the
morphology density relation (\eg\, Smith \etal\ 2005) indicate that
the pure luminosity evolution is not likely to be an appropriate model
for the evolution of the spheroidal component, although at the moment
there is not enough information to disentangle the dependency of the
demographics of spheroids on mass and environment. A measurement of
the evolution of the mass function of bulges and disks as a function
of environment is needed to make further progress.

\acknowledgements

LAM acknowledges financial support from the National Science and
Engineering Council of Canada.  TT acknowledges support from the NSF
through CAREER award NSF-0642621, by the Sloan Foundation and by the
Packard Foundation.  RSE acknowledges financial support from the Royal
Society. Some of the data presented herein were obtained at the W.M.\@
Keck Observatory, which is operated as a scientific partnership among
the California Institute of Technology, the University of California
and the National Aeronautics and Space Administration. The Observatory
was made possible by the generous financial support of the W.M.\@ Keck
Foundation.  This work is partly based on archival data from the {\it
Hubble Space Telescope}, obtained from the data archive at the Space
Telescope Institute, which is operated by the association of
Universities for Research in Astronomy, Inc.\@ for NASA under contract
NAS5-26555.

\end{document}